# Mask Responses for Single-Pixel Terahertz Imaging

Sven Augustin[1,2], Sven Frohmann[1,3], Peter Jung[4], Heinz-Wilhelm Hübers[1,2]

**Terahertz (THz) radiation meaning electromagnetic radiation in the range from 0.1 THz (3 mm) to 10 THz (30 µm) has the unique advantage of easily penetrating many obstructions while being non-hazardous to organic tissue since it is non-ionizing. A shortcoming of this domain is the limited availability of high-sensitivity detector arrays respective THz cameras with >1k pixels. To overcome the imaging limitations of the THz domain, compressive imaging in combination with an optically controllable THz spatial light modulator is a promising approach especially when used in a single-pixel imaging modality. The imaging fidelity, performance and speed of this approach depend crucially on the imaging patterns also called masks and their properties used in the imaging process. Therefore, in this paper, it is investigated how the image quality after reconstruction is specifically influenced by the different mask types and their properties in a compressive imaging modality. The evaluation derives specific guidelines for the use in THz single-pixel imaging setups respective THz single-pixel cameras employing optically controllable THz spatial light modulators. As an outlook, a method is suggested that builds upon the results presented in this text and extends the discussed advantages to an improved compressive imaging setting fit for stand-off security imaging applications aka single-pixel detector Body Scanners .**

Imaging at low THz frequencies provides unique information in particular for applications in security because at these frequencies a good compromise between penetration properties on one hand and spatial resolution on the other hand is possible. As a rule of thumb, the penetration properties greatly improve with decreasing THz frequency. However, the spatial resolution decreases at the same time. Another challenge is the lack of suitable multi-pixel THz cameras. Accordingly, the classical approach for security imaging at frequencies in the low THz region (below 1 THz) involves some form of mechanical scanning either with a single or a few detector elements [1], [2]. Due to the mechanical scanning the achievable frame rate is rather limited depending on the amount of necessary image pixels. To overcome the frame rate limitation the combination of a sensitive single-pixel detector with compressive imaging offers a potential solution.

Compressive imaging refers to an imaging process in which images can be acquired with fewer measurements than image pixels. One way to implement such a measurement scheme is the use of a spatial light modulator (SLM). When a SLM is combined with a single-pixel detector such an imaging setup is usually referred to as a single-pixel camera (SPC). The

[1] Humboldt Universität zu Berlin - Department of Physics Newtonstraße 15, 12489 Berlin

[2] German Aerospace Center Berlin - Adlershof, Rutherfordstraße 2, 12489 Berlin

[3] Technical University Berlin – Department of Optics and Atomic Physics, Straße des 17. Juni 135, 10623 Berlin

[4] Technical University Berlin – Communications and Information Theory Group Einsteinufer 25, 10587 Berlin

SPC concept (in this sense) for the visible part of the electromagnetic spectrum (VIS-SPC) was introduced in 2008 [3]. The general design of a SPC enables image acquisition with increased resolution [4], improved signal-to-noise ratio (SNR) [5] and larger depth of focus [6]. All these advantages can be achieved with a single-pixel detector even without mechanical scanning with the help of a SLM, which, in this case, acts as a dynamic aperture [7]. An image of the scene is formed from sequential measurements and an image reconstruction process that is based on solving an (underdetermined) system of equations (image reconstruction) [8]). The SPC concept is especially beneficial for cameras that work outside of the visible part of the electromagnetic spectrum (EM-spectrum) respective, for example for the THz domain.

Following the VIS-SPC concept, similar implementations in the THz domain have been realized and implementations for this domain are reported in [9], [10], [11], [12] and [13]. In these systems a THz-time-domain system [9], [13], a 0.35 THz multiplier source [11] or an incandescent light source [10] was used as a source of THz radiation. The general design of a SPC is similar for the VIS and the THz domain. It consists of a SLM, a radiation source and a single-pixel detector. A major difference between a VIS-SPC and a THz-SPC is the implementation of its SLM since THz-SLMs are not readily available. There are for example, THz-SLMs that are still in the early research stage, which are based on metamaterials as described in [12]. This THz-SLM approach addresses individual pixels electronically. While it makes such SLMs easy to control this approach has the shortcoming of providing only a very small amount of controllable pixels (so far <100 pixels were reported). Another approach that allows for more than 1500 individually controllable pixels is an optically controllable THz-SLM. In this approach a VIS-SLM is used to project spatial patterns (so called masks) onto a suitable semiconductor disc, the optical switch (OS). In the illuminated regions electrons are excited into the conduction band and the semiconductor (partially) changes from a semiconducting to a metallic state becoming less transmissive for THz radiation. This mechanism implies that the performance of an optically controllable THz-SLM is influenced by the quality and quantity of the OS's illumination and the OS material (the type and quality of the semiconductor) itself [14]. A detailed analysis of this subject, while extremely relevant for THz-SPCs, is beyond the scope of the analysis presented here. However, Figure 1 shows a generic block diagram of an optically controllable THz-SPC visualizing the concept just presented.

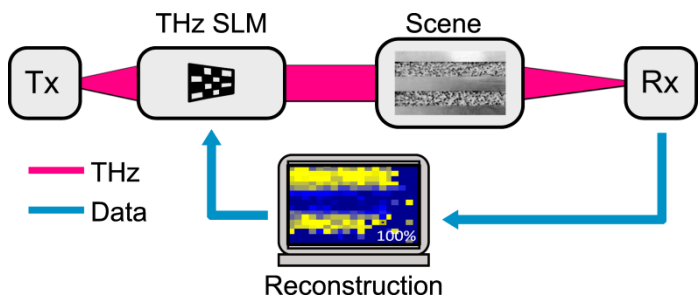

Figure 1: Scheme of the imaging process encountered in a THz-SPC. The beam coming from the THz source (Tx) is spatially modulated by a THz-SLM. The spatially modulated THz beam is directed to the scene of the camera and the radiation coming from the scene is detected using a single-pixel detector (Rx). With the knowledge of the spatial modulation patterns in connection with the measured responses an image of the scene can be reconstructed using a non-linear reconstruction algorithm[1].

---

[1] In this text, a non-negative least squares (NNLS) algorithm is used for reconstruction. The achievable reconstruction success respective the imaging fidelity depends crucially on the chosen reconstruction algorithm. NNLS was chosen due to its robustness properties in a compressive imaging modality. The performance of the NNLS algorithm is analyzed for binary Bernoulli masks here [15], [16]. The performance of NNLS for Hadamard and grayscale masks is still an open research topic.

3In this paper we investigate relevant mask properties used for SPC imaging. The text is thereby structured as follows. The next section "Experimental Setup and Procedure" introduces the setup in detail and describes the imaging procedure using a 0.35 THz SPC. In the section "Mask Types and THz Responses" the different investigated mask types are introduced and analyzed (mask responses). The following section "Image Reconstruction Results" presents reconstruction results for the different mask type measurements. For the analysis a single metallic edge is evaluated when fewer measurements than image pixels are acquired (compressive imaging modality). From this analysis conclusions and specific measurement guidelines for the use of specific mask types in an optically controllable 0.35 THz-SPC are derived. These conclusions are presented in the "Discussion" section and are applied to an application scenario where the imaging process is conducted from a distance of several centimeters (far-field). The text closes with a brief outlook how the given guidelines can be used to improve the performance of THz-SPCs in future stand-off security imaging experiments.

## Experimental Setup and Procedure

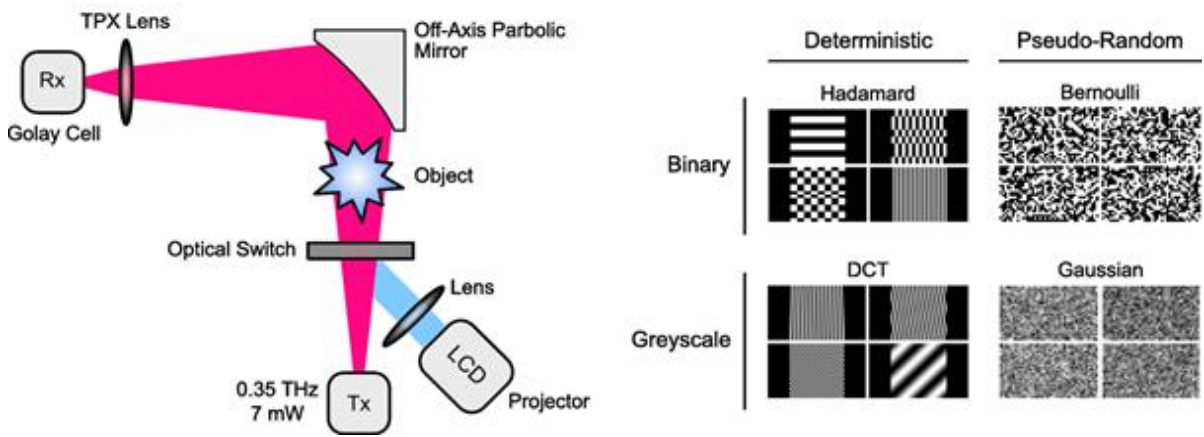

Figure 2: Single-pixel camera setup used for the measurements presented in the text (left-hand side). The THz-SPC in this case uses a commercially available projector as VIS-SLM and light source. The right-hand side shows the categorization with example masks investigated here.

The experimental setup is shown in Figure 2. The transmitter (Tx) is based on a Yttrium Iron Garnet oscillator, which is amplified and multiplied to 7mW of output power at a frequency of 0.35 THz. A horn antenna at the output of the Tx delivers a Gaussian-shaped beam, which impinges directly onto the OS made from passivated Silicon. This type of Silicon is especially sensitive to visible light (VIS), which enables the use of a commercial projector as VIS light source. To enable efficient coupling of the THz and the VIS radiation both optical paths are aligned at an angle of 45° with respect to each other. This enables a very compact design and increases the modulation but introduces geometrical distortions into the projected masks. An alternative design would be the use of a dichroic mirror e.g. made from glass coated with indium tin oxide. Such a mirror was tested but its use caused a reduction of modulation of the OS by a few percent. Therefore, the 45° geometry was used throughout the experiments. The THz radiation that passes through the OS is collected using a custom made off-axis parabolic mirror in combination with a lens made from Polymethylpentene (TPX).

This collecting optics was designed in such a way that the entire THz radiation passing through the OS is focused and detected by a single-pixel Golay cell detector.

The field-of-view (FoV) of this single-pixel camera was approximately 9.5 cm$^2$. It is mainly limited by the beam shape of the THz beam at the OS and not by the size of the OS itself. A projector was used to project spatial patterns (masks) onto the OS. In the illuminated regions the semiconductor (partially) changes from a semiconducting to a metallic state and becomes less transmissive for the 0.35 THz radiation impinging on the illuminated OS. The degree by which the THz transmission is changed upon illumination with a mask, i.e. the maximum modulation M, is evaluated using Equation (1).

$$M = 1 - \frac{I_{white}}{I_{black}} \qquad (1)$$

The maximum modulation (M) results from commanding an entirely white respective black mask to the THz-SPC. The resulting intensities are named $I_{white}$ and $I_{black}$. The modulation of spatially structured masks depends on the number of black respective white pixels in the masks and lies in between the responses of the entirely white respective black masks (see next section for details). With the setup described above, single-pixel imaging capability for 0.35 THz radiation is achieved with a maximum modulation M of approx. 15%.

## Mask Types and THz Responses

As already mentioned, the quality and type of different spatially structured masks directly influences the imaging performance in terms of achievable resolution, fidelity, imaging speed, compressibility, etc. Due to the implementation approach of the THz-SPC using a LCD based projector, a vast selection of masks can be used. To establish a systematic approach, the masks are classified here into four major categories. The first distinction is in regard to the mask type describing the fundamental structure. Here, two categories are used; the deterministic mask type and the pseudo-random mask type. All possible masks cannot be classified this way; since also masks exist that are a mixture of the deterministic and pseudo-random category. For both categories the masks can additionally be categorized as either binary or grayscale. In binary masks each pixel can be either fully transmissive (1) or not-transmissive (0), while in grayscale masks the transmission of each pixel can vary between 0 and 1. These four categories are investigated here using a prominent example for each category. This categorization including each investigated example is shown on the right-hand side of Figure 2. In the following, the main features of these masks are described.

**Hadamard masks**

They exhibit a periodic structure that increases in spatial frequency with increasing mask number (natural order). For the generation of the Hadamard masks used for the investigation presented here, the built-in functions of the MATLAB programming language were used. All masks are rescaled to the interval [0,255]. The periodic structure of the Hadamard masks gives a large correlation in the imaging process only with a limited number of spatial frequencies in the scene.

**Bernoulli masks**

Due to the pseudo-random structure of these masks each measurement gives information about large spatial frequency content of the scene. Thereby the Bernoulli masks are generated by choosing at random a value of 1 or 0 for each mask block. Each mask block is comprised of multiple SLM pixels. Again for the physical implementation the masks are rescaled to the interval [0,255].

**DCT masks**

The abbreviation DCT stands for Discrete Cosine Transform. These masks are grayscale, deterministic masks that contain multiple values in the interval $[-1,1]$. The DCT masks as deterministic, grayscale mask type example investigated here, are also generated with the help of built-in MATLAB functions and are rescaled to the interval [0,255]. The performance of DCT masks is especially interesting since the results can be related to the JPEG compression standard.

**Gaussian masks**

The last mask type investigated here uses Gaussian masks, which are masks where each mask block is drawn separately from a standard Gaussian distribution. The result is then rescaled to the interval [0,255].

Both grayscale mask type examples are of special theoretical interest since they allow the derivation of theoretical imaging limits (Gaussian masks) and come with fast reconstruction algorithms (DCT - non-compressive). Due to the fundamentally different nature of the four mask types, different THz-responses are expected when they are commanded to the THz-SPC. Each response is also influenced by the number of mask pixels with the same value (mask block size) and whether the structured part of the masks has to be quadratic or rectangular (orthogonal transforms). The responses, i.e. the intensity measured with the Golay detector for each commanded mask, of the four mask types are shown in Figure 3 as a function of the measurement number.

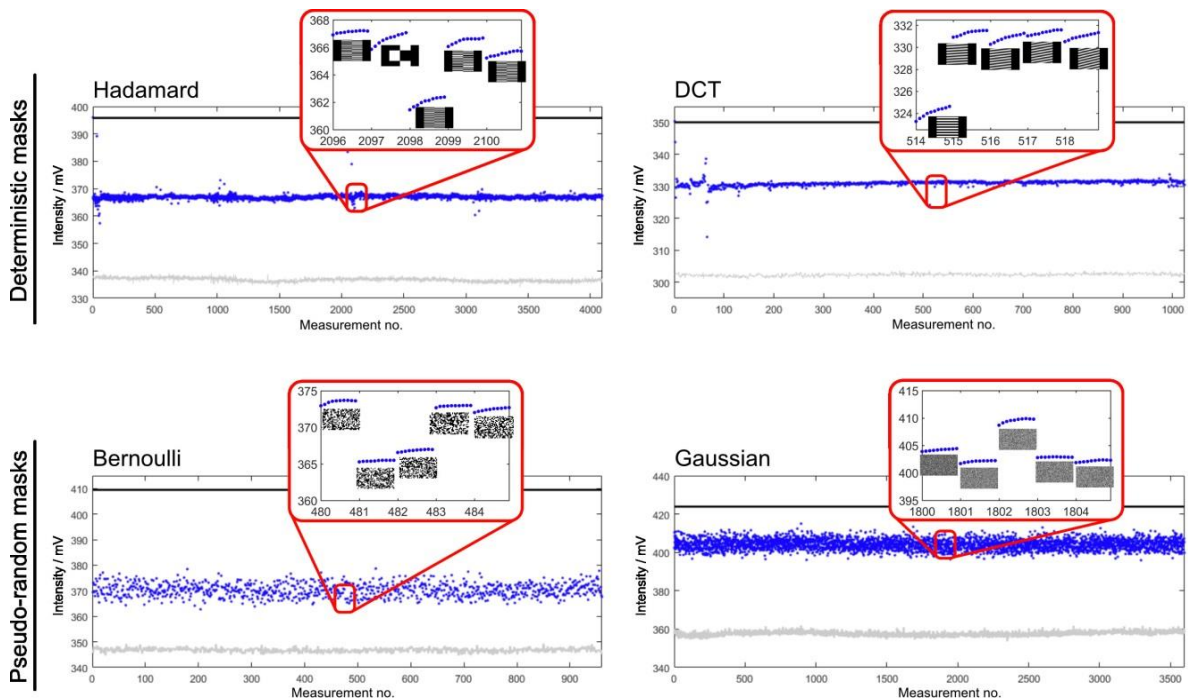

Figure 3: Mask responses of the different mask types investigated here.

Here, each measurement number corresponds to one particular mask of the respective category of masks. In each of the plots the upper black curve corresponds to the signal measured with the Golay detector when a completely black mask is commanded to the projector (i.e. no illumination of the semiconductor and maximum transmission of the THz beam) and the lower gray curve corresponds to the THz signal when a completely white mask is projected (i.e. the whole semiconductor is illuminated by the projector). The blue curve is the signal measured with the Golay detector when the semiconductor is illuminated with a spatially structured mask. In the insets some examples of commanded spatially structured masks and the corresponding THz signals are shown.

There are a number of general features which can be seen in Figure 3. First of all the intensity measured with the Hadamard and DCT masks strongly deviate from the average value only for a few specific masks. This corresponds to the fact that these masks have a strong correlation with the scene. The THz responses for the pseudo-random masks have a significantly larger average variation than those of the deterministic masks. However, unlike the Hadamard and DCT masks no value stands out. This fundamental difference in the measured responses indicates that the sequence of masks in a SPC imaging process is only important for deterministic masks and that the pseudo-random masks capture, with each measurement, large information content of the scene. If this hypothesis is true then a measurement with pseudo-random masks should be more robust against measurement error than deterministic masks. In addition, it also should provide measurements using pseudo-random masks a high capacity for compression. This hypothesis will be tested in the next section. The robustness and compressibility properties of the four mask type examples will be determined by analyzing the image reconstruction results of a single metallic edge in order to have comparable conditions for the different mask types. The mask type showing the most promising results in this evaluation is then used in a full compressive imaging modality with a metallic grid object that is placed a few centimeters behind the OS.

## Image Reconstruction Results

The reconstruction algorithm is used to solve the linearized imaging model stated in Equation (2)[2].

$$Y = \Phi \cdot X + N \qquad (2)$$

Here, Y denotes the vector of measured mask responses and X is the THz representation of the scene to be reconstructed. The matrix ɸ contains in each row the mask used for the respective measurement $Y_i$ and the vector N models an additive Gaussian noise contribution. As mentioned before, for the evaluation of the results Equation (2) is solved using a least-square algorithm with the additional constraint of non-negativity of the measured values $Y_i$ (Non-Negative Least Squares NNLS https://de.mathworks.com/matlabcentral/fileexchange/38003-nnls-non-negative-least-squares).

---

[2] The algorithm that is used to calculate an image from the measured THz responses has also significant influence on the reconstruction performance in terms of image fidelity, reconstruction speed, achievable compression factor, etc.

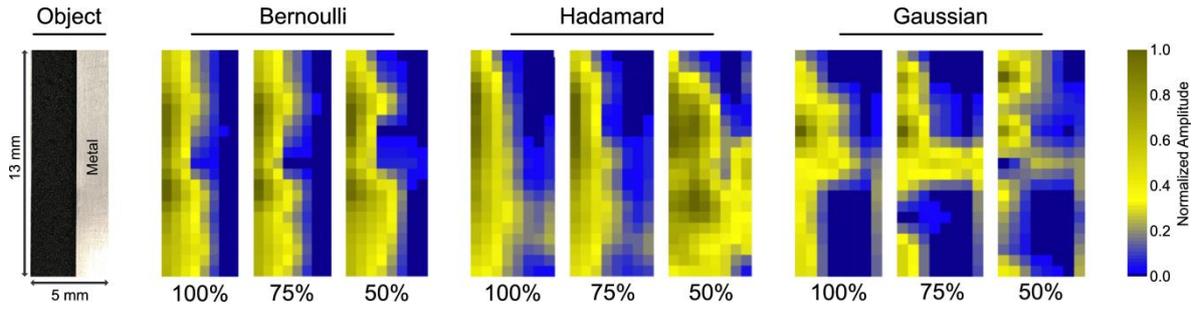

Figure 4: Reconstructions of a metallic edge using the NNLS algorithm for Bernoulli masks, Hadamard masks and Gaussian masks. A reconstruction using DCT masks was unsuccessful and is therefore not shown here. In each case the 100% case uses a number of measurements that is equal to the number of pixels.

Figure 4 shows the results of this comparison. Only Bernoulli and Hadamard masks are able to produce a reasonably good image of the single metallic edge shown in the left-hand side of Figure 4. The Gaussian masks produce a THz representation of the edge that shows already significant distortions in the case where the number of measurements is equal to the number of pixels (100%). In the compressive cases (number of measurements <100% randomly chosen values $Y_i$ are omitted for the reconstruction process (a detailed description of the signal processing method that was used to simulate the compressive imaging modality can be found in the Methods section). As can also be seen in Figure 4 only the Bernoulli masks show good compression results of the metallic edge even when only random 50% of the measured values are used for reconstruction.

According to this analysis, only the pseudo-random Bernoulli masks exhibit potential for compressive imaging. As evidence for this conclusion a metallic grid object imaged with the 0.35 THz SPC in a compressive modality is shown in Figure 5.

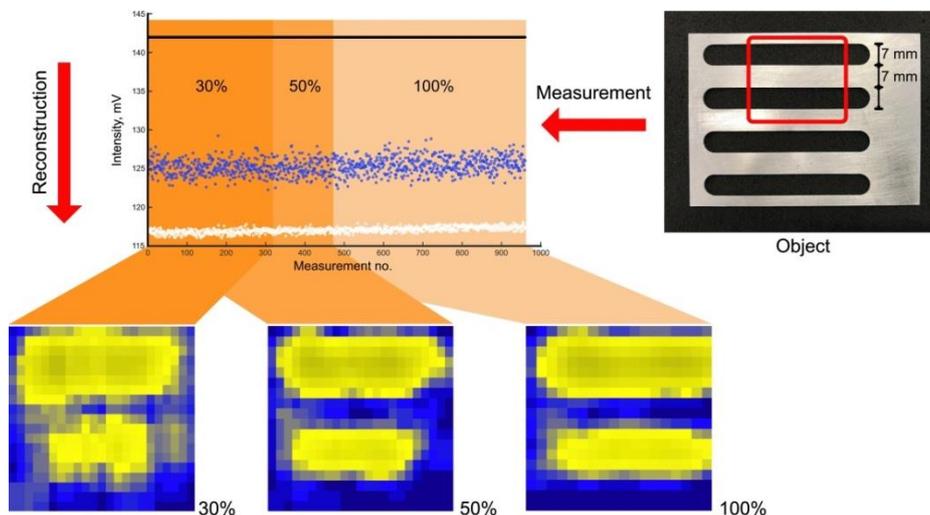

Figure 5: Imaged metal grid scene target using Bernoulli masks. Shown are the image reconstructions with a reduced number of considered measurements (100 % - 30 % when Bernoulli masks are commanded to the SPC. The scene target was a metallic grid. The target has 7mm wide metallic bars and 7mm wide open spaces (binary target - see photo for details).

Even with only 30% of all masks the shape of the imaged object is recognizable albeit with smaller SNR. The hypothesis that Bernoulli masks are well suited for a compressive imaging modality is clearly supported by the result shown in Figure 5. Although the SNR is smaller for the 30% compression case and the metal edges are smeared out the structure is still recognizable. The result might even be improved using a sparsifying transformation [19]. Such a transformation could not be applied to the results presented here since sparsifying transformations for 0.35 THz are still an open research subject.

## Discussion

To summarize and discuss the results of the investigation just presented, specific guidelines for SPC measurements can be derived.

The Hadamard mask measurements provide a large modulation for specific masks but in a compressive imaging modality, where a random part of the measurements are not considered for reconstruction, their application was found to be not optimal. The compressive modality is often encountered in 0.35 THz single-pixel imaging applications since the output power of the Tx itself is drifting during a measurement, the coherent nature of the radiation causes interference effects or even standing waves. This is especially true in cases where the scene is several centimeters or even meters away from the optical switch (security imaging applications). Mask measurements acquired under the aforementioned conditions lead to measurement errors (transmission errors) that should be omitted for the reconstruction. This approach will automatically lead to a compressive imaging modality. Due to this reason the use of pseudo-random masks is supported here. The performance of pseudo-random masks under compressive conditions can even be improved with a SNR like quantity, which would provide an indicator whether a SPC measurement was successful even without image reconstruction. This quantity should even give a quantitative measure for the decision which mask measurements can be considered for reconstruction

As shown by the investigation presented here only pseudo-random mask type measurements give sufficient compressibility in the pixel domain. The compressibility of pseudo-random masks coincides with a robustness property against transmission /measurement error [17]. It is assumed that this property stems from the ability of pseudo-random masks to acquire information content of the entire scene with each measurement. This means that the image content is not diminished with increasing undersampling factors only the SNR of the reconstructed images. Compared with a traditional (software) Raster scan the increase in SNR and robustness of pseudo-random mask measurements theoretically outweigh the Raster scan approach by several orders of magnitude. The exact figure depends on the number of image pixels, details of the SPC implementation as well as the information content of the scene target. As a rule of thumb, the SNR increase is of the order N/2 where N is the number of image pixels.

However, the question "what are the best masks?" can only be answered knowing the specific imaging task at hand. Additionally, the identified beneficial properties of pseudo-random masks can be significantly improved with the help of sparse domains, in other words, transformations of the scene that significantly improve its sparsity. For 0.35 THz single-pixel imaging such domains are still an open research topic[3].

The analysis also suggests that the use of grayscale masks is not possible for 0.35 THz-SPCs without a suitable calibration of the camera. The calibration would also be beneficial in terms of reducing measurement overhead and towards imaging grayscale targets. First approaches on the subject are mentioned here [21] and may prove helpful. As long as no calibration procedure for 0.35 THz-SPCs exists, imaging is confined to binary masks and binary objects. As stated before, the robustness and compressibility properties of pseudo-random masks (Bernoulli masks) are very beneficial for the imaging process. Still, the reconstruction approaches are still lacking reconstruction speed. When real-time reconstruction speed is necessary Hadamard masks are still a good choice.

---

[3] Shearlets may be an adequate way to find a solution for this task see [18] due to the cartoon-like nature of images in this part of the EM-spectrum.

## Outlook

So far the investigation excluded the dimension of mask block size. A larger block size increases measurement SNR but on the other hand decreases the achievable spatial resolution in the reconstructed images. Additionally, the effect due to the coherent nature of the THz radiation was not considered and probably plays a role in relation to the mask block size. Since this may be a limiting factor for image fidelity and image resolution, it will be considered in future experiments. The goal of these future experiments is the successful application of the THz-SPC concept for real world (grayscale) targets [22]. For this application a reflection modality even allows security imaging applications, which is especially useful when combined with a radar approach [23].

## Methods

In order to simulate the compressive imaging modality for all image acquisitions with the THz-SPC, the number of measured structured masks was equal to the number of pixels. From these measurements a portion commensurate to the undersampling ratio/compression ratio was selected at random. Only these measurements were considered in the reconstruction process. Additionally, for each structured mask two unstructured masks (black, white) were measured in order to account for power drifts during the measurements. This drift correction was implemented using the assumption that the value for the measured black masks is constant throughout the entire THz-SPC imaging process. This assumption gives a scaling factor for each structured mask.

## References

[1] E. Heinz, T. May, D. Born, G. Zieger, A. Brömel, S. Anders, V. Zakosarenko, T. Krause, A. Krüger, M. Schulz, F. Bauer, H.-G. Meyer, N. A. Salmon, and E. L. Jacobs, *"Development of passive submillimeter-wave video imaging systems for security applications,"* in SPIE Security + Defence, ser. SPIE Proceedings. SPIE, p. 854402, 2012.

[2] K. B. Cooper, R. J. Dengler, N. Llombart, B. Thomas, G. Chattopadhyay, and P. H. Siegel, *"Thz imaging radar for standoff personnel screening,"* IEEE Transactions on Terahertz Science and Technology, vol. 1, no. 1, pp. 169–182, 2011.

[3] M. F. Duarte, M. A. Davenport, D. Takhar, J. N. Laska, Ting Sun, K. F. Kelly, and R. G. Baraniuk, *"Single-pixel imaging via compressive sampling,"* IEEE Signal Processing Magazine, 25(2):83–91," 2008.

[4] David B. Phillips, Ming-Jie Sun, Jonathan M. Taylor, Matthew P. Edgar, Stephen M. Barnett, Graham M. Gibson, and Miles J. Padgett, „*Adaptive foveated single-pixel imaging with dynamic supersampling,*" Science advances, 3(4):e1601782, 2017.

[5] Ming-Jie Sun, Matthew P. Edgar, David B. Phillips, Graham M. Gibson, and Miles J. Padgett, „*Improving the signal-to-noise ratio of single-pixel imaging using digital microscanning,*" Optics Express, 24(10):10476–10485, 2016.

[6] Ryoichi Horisaki, Hiroaki Matsui, Riki Egami, and Jun Tanida, „*Single-pixel compressive diffractive imaging,*" Applied Optics, 56(5):1353, 2017.

[7] A. Zomet and S. K. Nayar, „Lensless imaging with a controllable aperture," In Cordelia Schmid, Stefano Soatto, and Carlo Tomasi, editors, 2006 IEEE Computer Society Conference on Computer Vision and Pattern Recognition workshops, pages 339–346, Piscataway, NJ, 2006.

[8] Marcin Kowalski, Marek Piszczek, and Mieczyslaw Szustakowski, „*Test environment for image synthesis of a single pixel camera,*" SPIE Proceedings, page 85421G. 2012.

[9] Stefan Busch, Benedikt Scherger, Maik Scheller, and Martin Koch, „*Optically controlled terahertz beam steering and imaging,*" Optics Letters, 37(8):1391–1393, 2012.

[10] David Shrekenhamer, Claire M.Watts, and Willie J.Padilla, „*Terahertz single pixel imaging with an optically controlled dynamic spatial light modulator,*" Optics Express, 21(10):12507–12518, 2013.

[11] S. Augustin, J. Hieronymus, P. Jung, and H.-W. Hübers, „Compressed sensing in a fully non-mechanical 350 ghz imaging setting," Journal of Infrared, Millimeter, and Terahertz Waves, 36(5):496–512, 2015.

[12] Claire M. Watts, David Shrekenhamer, John Montoya, Guy Lipworth, John Hunt, Timothy Sleasman, Sanjay Krishna, David R. Smith, and Willie J. Padilla, „*Terahertz compressive imaging with metamaterial spatial light modulators,*" Nature Photonics, 8(8):605–609, 2014.

[13] Zhenwei Xie, Xinke Wang, Jiasheng Ye, Shengfei Feng, Wenfeng Sun, Tahsin Akalin, and Yan Zhang, „*Spatial terahertz modulator,*" Scientific Reports,3(1):97, 2013.

[14] Tom F. Gallacher, Rune Sondena, Duncan A. Robertson, and Graham M. Smith. „*Optical modulation of millimeter-wave beams using a semiconductor substrate,*" IEEE Transactions on Microwave Theory and Techniques, 60(7):2301–2309, 2012.

[15] Richard Kueng and Peter Jung, „*Robust nonnegative sparse recovery and 0/1-bernoulli measurements,* " In 2016 IEEE Information Theory Workshop (ITW), pages 260–264, Piscataway, NJ, 2016.

[16] Ukash Nakarmi and Nazanin Rahnavard. Bcs: „*Compressive sensing for binary sparse signals,*" In IEEE Military Communications Conference, 2012 - MILCOM 2012, pages 1–5, 2012.

[17] S. Augustin and H.-W. Hübers, „*Understanding mask switching for thz compressive imaging,*" In 2016 41st International Conference on Infrared, Millimeter, and Terahertz Waves (IRMMW-THz), pages 1–2, 2016.

[18] J. Hieronymus, S. Augustin, and H.-W. Hübers, „*Characterization of a thz slm and its application for improved high resolution thz imaging,*" In 2015 40th International Conference on Infrared, Millimeter, and Terahertz waves (IRMMW-THz), pages 1–2, 2015.

[19] Vishal M. Patel and Rama Chellappa. „*Sparse representations, compressive sensing and dictionaries for pattern recognition,*" In Tieniu Tan, editor,First Asian Conference on Pattern Recognition (ACPR), 2011, pages 325–329, 2011.

[20] Gitta Kutyniok and Demetrio Labate, editors, „*Shearlets: Multiscale analysis for multivariate data,*" Applied and numerical harmonic analysis, Birkhäuser / Springer, Boston, Mass., 2012.

[21] R. Gribonval, G. Chardon, and L. Daudet. „*Blind calibration for compressed sensing by convex optimization,*" In 2012 IEEE International Conference on Acoustics, Speech and Signal Processing (ICASSP), pages 2713–2716, 2012.

[22] S. Augustin, S. Frohmann, P. Jung, T. Schulze, and H. W. Hübers. „*Towards real-world Thz imaging with optically controllable single-pixel cameras,*" In Optical Teraherz Science and Technology (OTST), 2017.

[23] B. Fürsich, S. Augustin, R. Bamler, X. Zhu, and H. W. Hübers. „*Towards a Thz stand-off single-pixel camera using compressed sensing,*" In 4th CoSeRa workshop, 2016.

## Acknowledgements

The authors gratefully acknowledge support by the German Research Foundation (DFG) under the grants HU848/7 and JU2795/3 in the scope of the priority program SPP1798 Compressed Sensing in Information Processing (CoSIP).

Contributions

S. A. designed and built the setup, conducted the experiments and was involved in the image reconstruction. S. F. contributed to the discussion of the results and the image layout process. P. J. contributed to the image reconstruction and the image analysis. H.-W. H. contributed to the design of the setup and the analysis of the data. All authors were involved in the process of writing the manuscript.

Additional Information

**Competing financial interests:** The authors declare no competing financial interests.